\documentclass[sn-mathphys,Numbered]{sn-jnl}

\usepackage{graphicx}%
\usepackage{multirow}%
\usepackage{amsmath,amssymb,amsfonts}%
\usepackage{amsthm}%
\usepackage{mathrsfs}%
\usepackage[title]{appendix}%
\usepackage{xcolor}%
\usepackage{textcomp}%
\usepackage{manyfoot}%
\usepackage{booktabs}%
\usepackage{algorithm}%
\usepackage{algorithmicx}%
\usepackage{algpseudocode}%
\usepackage{listings}%
\usepackage{natbib} 
\theoremstyle{thmstyleone}%
\theoremstyle{thmstyletwo}%

\theoremstyle{thmstylethree}%

\begin{document}

\title[Article Title]{Passive None-line-of-sight imaging with arbitrary scene condition and detection pattern}


\author[1,2,3]{\fnm{Yunting} \sur{Gui}}\email{guiyunting@outlook.com}
\equalcont{These authors contributed equally to this work.}

\author[1,2]{\fnm{Xueming} \sur{Xiao}}\email{alexcapshow@gmail.com}
\equalcont{These authors contributed equally to this work.}

\author*[1,2,3]{\fnm{Yuegang} \sur{Fu}}\email{fuyg@cust.edu.cn}

\author[4]{\fnm{Meibao} \sur{Yao}}\email{meibaoyao@jlu.edu.cn}

\affil*[1]{\orgdiv{School of Optoelectronic Engineering}, \orgname{Changchun University of Science and Technology}, \orgaddress{\city{Changchun}, \postcode{130000}, \country{China}}}

\affil[2]{\orgdiv{CVIR lab}, \orgname{Changchun University of Science and Technology}, \orgaddress{\city{Changchun}, \postcode{130000}, \country{China}}}

\affil[3]{\orgdiv{Key Lab of Opto-electronic Measurement and Optical Information Transmission Technology, Ministry of Education}, \orgname{Changchun University of Science and Technology}, \orgaddress{\city{Changchun}, \postcode{130000}, \country{China}}}

\affil[4]{\orgdiv{School of Artificial Intelligence}, \orgname{Jilin University}, \orgaddress{\city{Changchun}, \postcode{130000}, \country{China}}}


\abstract{Passive Non-Line-of-Sight (NLOS) imaging aims to reconstruct objects that cannot be seen in line without using external controllable light sources. It can be widely used in areas like counter-terrorism, urban warfare, and autonomous-driving. Existing methods for passive NLOS imaging typically require extensive prior information and significant computational resources to establish the light transport matrices and train the neural networks. These constraints pose challenges for transitioning models to different NLOS scenarios. Thus, we declare that the pressing issue in passive NLOS imaging currently lies in whether it is possible to estimate the light transport matrices that correspond to relay surfaces and scenes, as well as the specific distribution of targets with only few shot. In this work, we hypothesized and mathematically proved the existence of a high-dimensional manifold, where the structural information of obscured targets is minimally disrupted. Based on this, we propose a universal framework named High-Dimensional Projection Selection (HDPS), which can output its projection onto corresponding low-dimensional surfaces. HDPS can be applied to most mature network architectures and allows estimating the distribution of targets and light spots obtained by a camera with minimal prior data. As we demonstrate experimentally, our framework, even applied to the most basic network structures, can achieve better results with significantly smaller amounts of prior data. Thereby, our approach enables passive NLOS scenarios to reconstruct targets with limited prior data and computational resources.}


\maketitle
\section{Introduction}\label{sec1}
The capacity to perceive objects that are unseen from the direct line of sight presents an interesting yet, extremely challenging topic in fields such as counter terrorism, urban warfare, and autonomous driving. By analyzing the scattered light on the Line-Of-Sight(LOS) wall, None-Line-Of-Sight(NLOS) imaging techniques~\cite{9156511} provide a potential way to reconstruct the image of hidden objects or objects that cannot be photographed directly. 

In recent years, the detection of concealed objects under NLOS conditions has garnered significant attention. Based on the utilization of controllable light sources, it can be categorized into active and passive imaging. In the context of the active NLOS solution, additional information such as time-of-flight (TOF) or structured light is used to reconstruct hidden objects~\cite{nam2021low}. Through triple reflection, the high-resolution time-resolved detector could capture the light information, making it easy to reconstruct the object. Another type of methods forewent TOF and opted optical phase encoding instead, reconstructing objects by deciphering information from two virtual light phase~\cite{liu2023non}. In terms of passive imaging, \citet{torralba2014accidental} were among the early pioneers to observe that elements within the environment can unintentionally function as cameras, simplifying the task of resolving concealed structures. They specifically exploited the characteristic that numerous everyday objects exhibit behavior akin to pinspecks or pinholes. In 2019, \citet{saunders2020multi}employed pinspeck occluders to unveil hidden details within 2D scenes. Additionally, \citet{yedidia2019using}used unknown occluders to recover hidden scenes. Moreover, in 2021, \citet{geng2021passive} reconstructed scene with the help of optimal transportation. However, with the poor information that uncontrollable light sources brought, research on passive NLOS imaging encountered a barrier when only a few shots is given as prior data.

\begin{figure}[htbp]%
\centering
\includegraphics[width=0.9\textwidth]{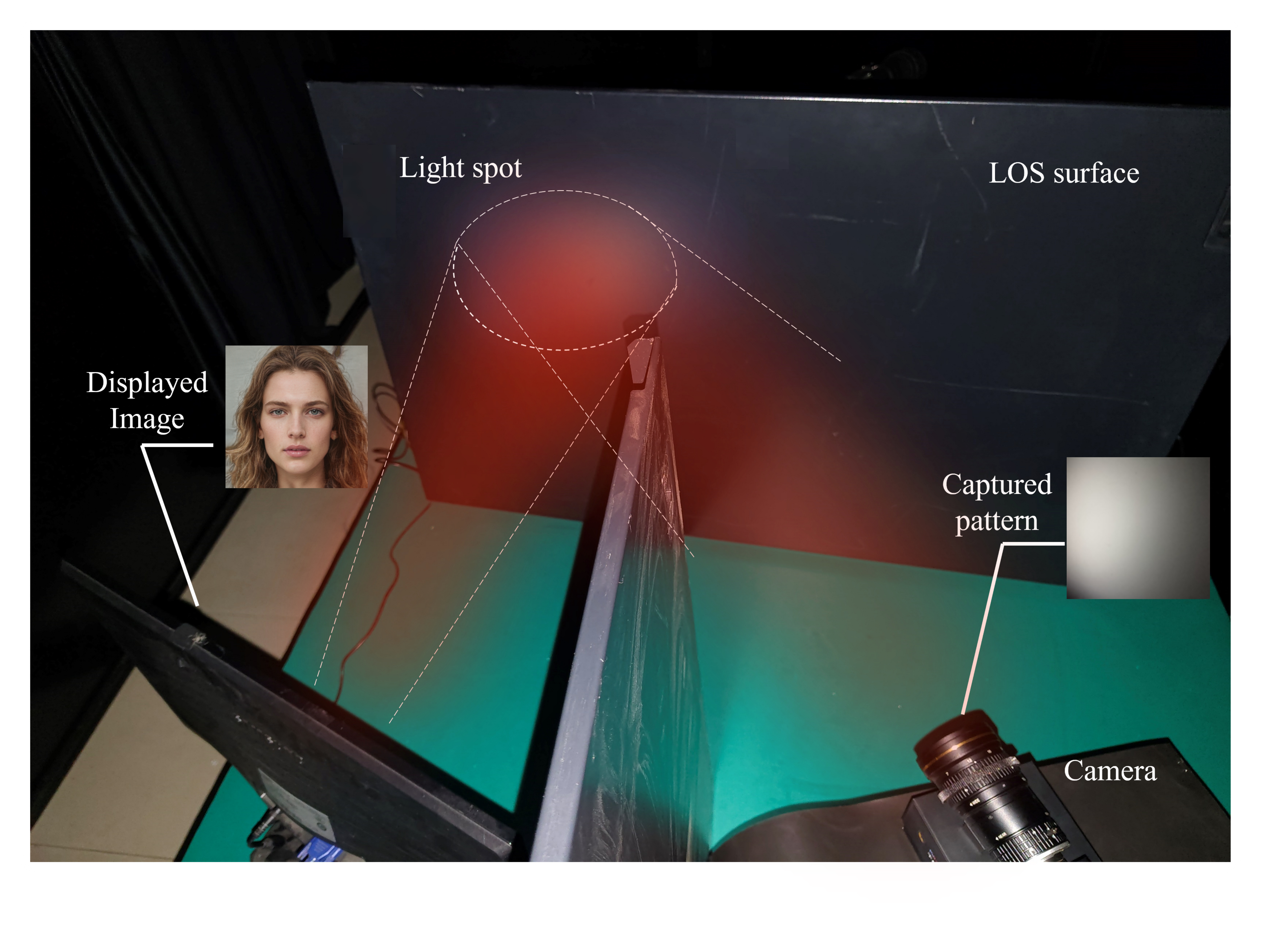}
            \caption{\textbf{Passive NLOS imaging experimental platform}. The experiment is conducted in a darkroom with a black metal panel as the unknown LOS surface and a computer screen as the light source. A camera is placed on the other side of the baffle plate. With each three seconds, images are randomly displayed on the screen. The camera will capture the scattering pattern reflected by the black panel simultaneously.}\label{fig1}
\end{figure}

Recent NLOS imaging methods in the passive domain have attempted to utilize basic deep-learning networks such as U-Net or structures based on transformers. With the assistance of physical information, these reconstruction methods have produced coarse results with blurred reconstructed images. However, these methods came with certain limitations, such as the necessity for a substantial database to serve as prior information or the requirement for a specific camera shooting angle. To address this issue, in this work, we originally developed a novel theoretical framework for general passive NLOS imaging. By establishing the correlation model between images and scattered light in high dimensional manifold space, our method can effectively eliminate the influence of optical scattering by wall, making it possible for image reconstruction from angle-free shooting. As shown in Fig.\ref{fig1}, we use a black metal panel as the LOS surface to simulate the unknown surface in the real world, where our method could reconstruct the scene shown in the display successfully. Furthermore, benefiting from the proposed high-dimensional manifold, reliable results could be achieved without the complex networks and even under small datasets as prior information.  

\section{Method}\label{sec2}
\subsection{The NLOS physical model}
The reconstruction issue in NLOS scenarios demands the retracing of the process whereby light is reflected , scattered, and other various optical phenomena by air or walls. Without occlusion, when the target is a point light source (approaching infinitesimally small in space), the intensity distribution of generated image follows the point spread function(PSF). However, PSF takes no effect under the NLOS condition. To this end, we proposed a solid scattering function(SSF), validated by experimental verification. It estimates the degree of diffusion response of light that undergoes scattering and refraction through a wall or other rough solid surface. Similarly with PSF, the spatial-based optical transfer function of imaging systems, SSF treats that the light scattered by the LOS (Line-of-Sight) wall still carried equivalent or similar structural information within a certain manifold during propagation. We posit that the light information naturally emitted by the occluded object can be regarded as a beam of structured light that has undergone encoding in an unknown high-dimensional space, carrying information about the structure of themselves. Under this assumption, a structured light beam group, after being scattered by air and the LOS wall, can maintain its encoded pattern in a high-dimensional manifold space without disruption. Therefore, the speckle pattern in the two-dimensional space will exhibit consistency with the target domain in a higher-dimensional space. In other words, the speckled image is the projection of the high-dimensional field onto a two-dimensional space. Based on that, by the decoding process of a neural network, it is possible to find the target manifold, specifically the inverse function of the SSF.
\begin{figure}[htbp]%
\centering
\includegraphics[width=0.8\textwidth]{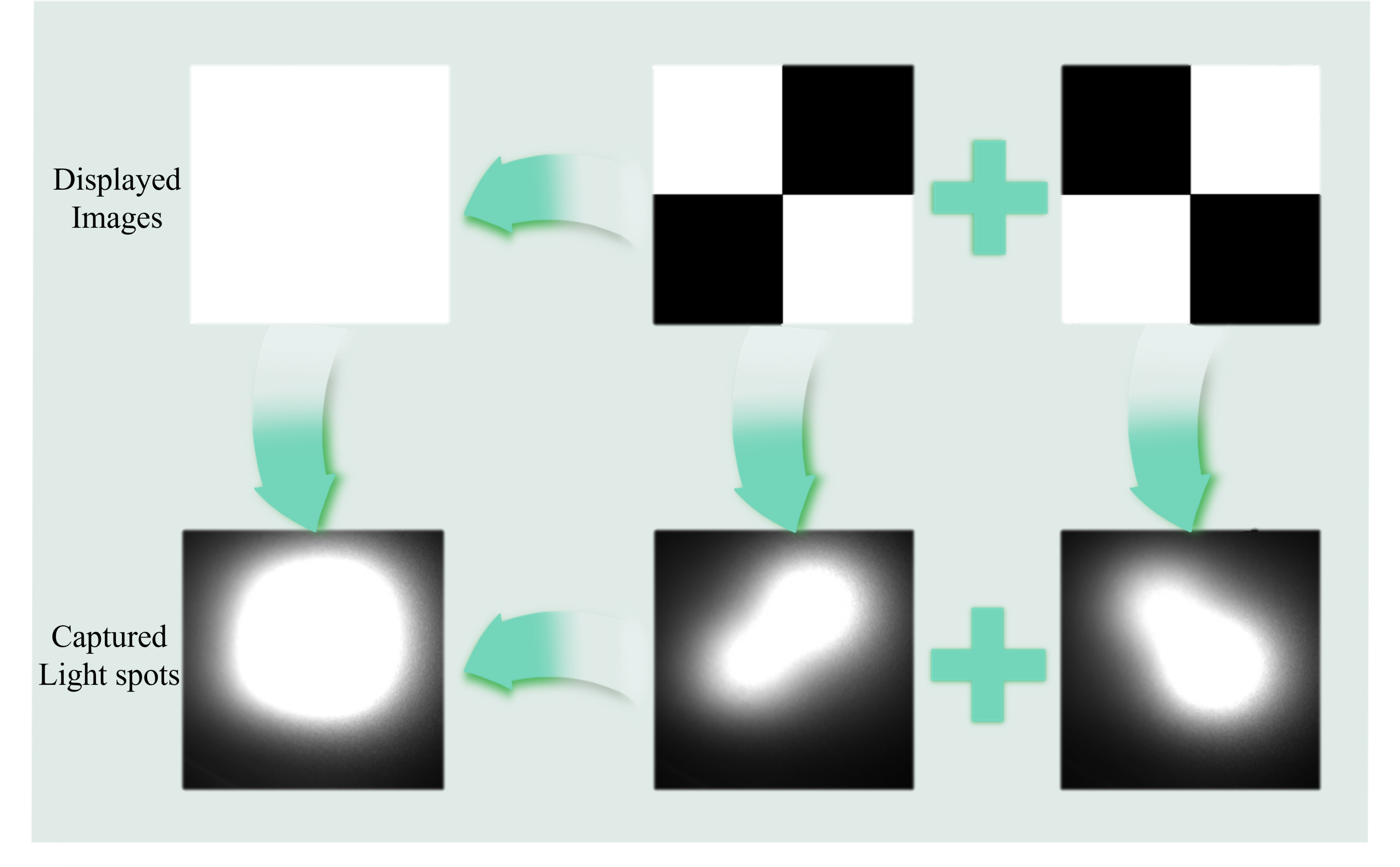}
            \caption{\textbf{The light field in NLOS imaging systems exhibits linear superposition.} Three ground-truth images are exhibited on the top line, the right-most two representing decompositions of the first left-most image. Subsequent to undergoing scattering within the non-line-of-sight (NLOS) system, the camera captures the ensuing two optical patterns. These two patterns can be rigorously linearly superimposed to constitute the first pattern, thereby manifesting the phenomenon of linear superposition.}\label{fig2}
\end{figure}

In the realm of optical systems, incoherent imaging systems manifest linear characteristics, as shown in Fig.\ref{fig2}. Therefore, the responses elicited by distinct optical field components introduced into the system are independent and additive: 
\begin{equation}
\text{Img}(\alpha X_1 + \beta X_2) = \alpha \text{Img}(X_1) + \beta \text{Img}(X_2).
\label{eq1}
\end{equation}

Same with PSF, SSF also exists as a one-to-one correspondence between the object plane and the image plane and possess the spatial invariance:
\begin{equation}
I(x,y)=\iint \text{SSF}(o(u,v))\,du\,dv.
\label{eq2}
\end{equation}
where I(x,y) is the scattering pattern and o(u,v) is the displayed image. We aim to design a generated framework that uses the neural network $f_\theta$ to reconstruct o(u,v) from each input I(x,y):
\begin{equation}
o(u,v) = f_\theta (I(x,y)).
\label{eq3}
\end{equation}
The inherent characteristics of the neural network required it must be continuous, owing to the presence of the back-propagation step. With the one-to-one correspondence between I(x,y) and o(u,v), we can easily prove the existence of SSF.

\subsection{The construction of high-dimensional manifold spaces}
The high-dimensional manifold contains various information from both the target and input manifolds. To reconstruct the NLOS scenes, we aspire to find such high-dimensional manifold space, and a pathway on it to connect all associated low-dimensional projections, simultaneously. As such, the relationships between low-dimensional manifolds, i.e. the target and input, can be performed by the reconstructed high-dimensional manifold as the medium. However, due to the inevitable information loss during projection, it remains an extremely challenging topic to establish bijective relationships between different projections in high-dimensional space. 

To this end, we define the space formed by all target images $o(u,v)$ as the target domain ${x}$, while the scattering pattern $I(x,y)$ space as the source domain ${z}$. Since each information in ${x}$ or ${z}$ is an image, we can directly utilize the Euclidean distance of intensity information to measure the distance of the metric $\rho$. It can be inferred that {x} and {z} satisfy the properties of positive definite, symmetry, and triangle inequality with respect to the metric $\rho$, as shown in Eq.\ref{eq4}:
\begin{equation}
\begin{aligned}
& max(\delta_i,\delta_j) < |x_i-x_j|, A_i\cap A_j = \emptyset\\
& min(\delta_i,\delta_j) > |x_i-x_j|, A_I \cap A_j = \emptyset.
\end{aligned}
\label{eq4}
\end{equation}
Therefore, the pair of (x, $\rho$) and (z, $\rho$) fulfills the first-countable axiom\footnote{For each element$\{x_i\}$ in $\{x\}$,$\exists$ $|x_i - a_i| < \delta_i$, definite set $A_i(a_i,\delta_i)=\{x_i|(x_i-a_i)<\delta_i\}$, so does $z_j$ and $B_j(b_j,\delta_j)$}.

According to the theory of point-set topology, given the sets $X=\bigcup_{i=1}^{n} A_i$ and $Z=\bigcup_{j=1}^{n} B_j$, denote $M$ as a family of subsets of $X$, and $X,\emptyset \in M$, the union of any arbitrary number of elements in $M$ still belongs to $M$. Eq.\ref{eq4} illustrates that M is the topology on $X$, while similar to the M, we find that $\exists N$ is the topology on $Z$. 

Thus, both $(X,M)$ and $(Z,N)$ are topological spaces. We further assumed the existence of high-dimensional smooth manifolds $fx$ and $fz$ and applied the neural network to fit the data. With a one-to-one correspondence between the single input $x$ ($x \in \{X\}$) and the single output $f_x$ ($f_x \in \{fx\}$), \textbf{there should be a bijective relationship, as proof shows, between the manifolds $fx$ and $X$}. By analogy, there will also be a bijective relationship between $fz$ and $Z$. Hence, due to the linearity of incoherent imaging systems as demonstrated in Sec.2.1, there must exist a one-to-one correspondence between the target sample and the captured input.
\begin{proof}
    $f(x)$ = $f_x$,
    $\forall f_x$ in $fx$ is the image of a certain element in $X$, $f$ can establish a surjection from $X$ to $fx$; for $\forall \varepsilon$ there must exist $\varepsilon < \delta_i$, $f_x - x\ge\varepsilon$, while $x_1 \neq x_2 \iff f_{x1} \neq f_{x2}$. It can establish an injective function from $X$ to $fx$.
\end{proof}

Due to the fitting properties of neural networks, it can be considered that both $f$ and its inverse function possess continuity. Therefore, it may be regarded that $fx$ and $X$ have homeomorphic properties, and similarly, there is a homeomorphic relationship between $fz$ and $Z$. In the context of NLOS imaging, the task of reconstructing $X$ from $Z$ can be extended to the process of mapping from the high-dimensional smooth manifold space $fz$ to $fx$. The way of extending information across manifolds via operators can effectively ensure the mapping of information between them.

\subsection{HDPS method}
As commonly acknowledged in the realm of image reconstruction research, there is a close interconnection between image generation and image reconstruction. The objective of image reconstruction is to extract specific information from a given input distribution that is either damaged or carries the target image information\cite{zhu2023harnessing}, establishing a direct correspondence with the target sample, thereby facilitating the reconstruction of the target image. On the other hand, image generation entails extraction from a given data distribution to create synthetic samples\cite{liu2023deep}. In the field of image generation, there exists a plethora of well-known algorithms encompassing foundational techniques such as GANs (\citet{goodfellow2014generative}), pixel autoregressive methods (\citet{van2017neural}), diffusion models (\citet{sohl2015deep}), VAE (\citet{kingma2013auto}) and some more recent, like consistency models (\citet{song2023consistency}) and Bayesian flow networks (\citet{graves2023bayesian}). Moreover, from these foundational algorithmic models for image generation, numerous algorithms for image reconstruction have emerged. Examples include pix2pix-GAN(\citet{zhu2017unpaired} )for image translation, latent diffusion models, and Nerf for three-dimensional scene reconstruction.

\begin{figure}[htbp]%
\centering
\includegraphics[width=1.0\textwidth]{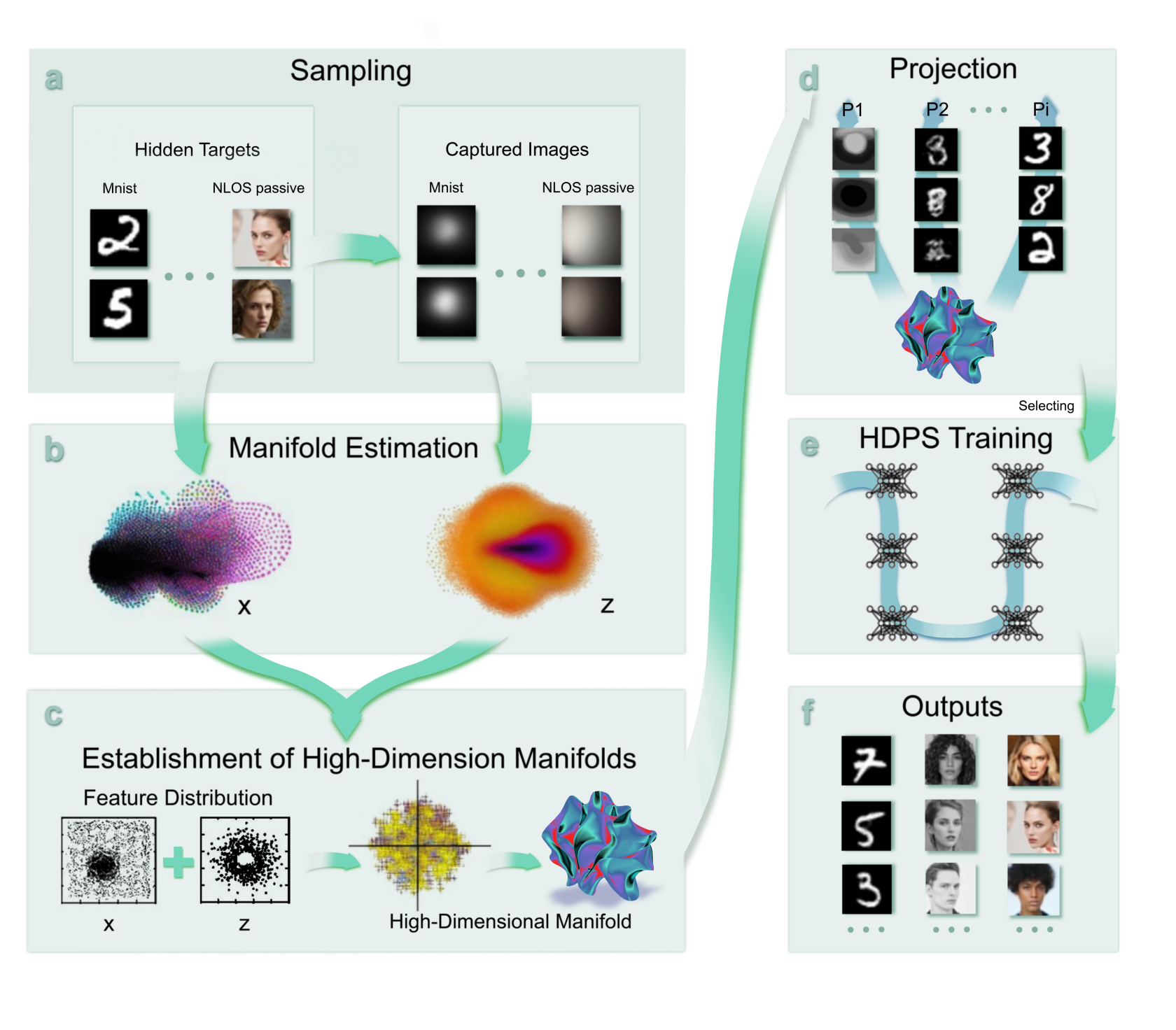}
            \caption{\textbf{Process of the HDPS method.} \textbf{a} describes the process when the hidden targets and light spots, as the samples of manifolds, input to the model. \textbf{b} is the process that the model extract the features and estimate the manifolds. \textbf{c} gives a brief introduction about how HDPS interact the manifold information and establish the high-dimensional manifold. \textbf{d} shows the projections that projected by high-dimensional manifold. \textbf{e} exhibits the training of HDPS model. \textbf{f} is the output of the model.}\label{fig3}
\end{figure}

The aim of these models is to directly provide outputs corresponding to target distributions through certain information extraction techniques and mapping methods applied to the input information. However, in cases where the degree of image corruption is significant or when the distribution of input images diverges highly from that of the target images, existing models struggle to produce accurate results with limited data samples or computational resources. Due to factors such as sparse data, some conventional models may fail to converge or become trapped in local minima, rendering them incapable of producing meaningful outcomes. To achieve the ambitious goal of enabling a model to operate with minimal data samples, accommodating various levels of damage and intricate optical transformations in the input, such as lens impairment, focus failure, or encountering images from Non-Line-of-Sight (NLOS) scenarios with entirely unknown light sources and material roughness on the Line-of-Sight (LOS) plane, it is essential to discover a high-dimensional information manifold encompassing both input and target domains. This high-dimensional manifold has infinite projections for which we only need to find an image manifold closely resembling the target domain.  As an initial step towards this endeavor, this work investigates a recovery algorithm, termed High-Dimensional Projection Selection (HDPS), which is an optimization principle that can be applied to most neural networks for image reconstruction. Fig.\ref{fig3} shows how HDPS worked to reconstruct the hidden targets.

As a few-shot learning framework, our HDPS model tries to estimate the distributions of target and input using minimal data. For such estimation, when the moments of all orders of the data tend to be similar, it will be feasible to extrapolate the global distribution utilizing a smaller dataset. Capitalizing on this trait, one can establish the global distribution information by refining the estimation of sample distributions through gradient update operators.

Meanwhile, by estimating the distributions of two low-dimensional manifolds and analyzing their inherent properties, the high-dimensional manifold can be constructed. Such high-dimensional manifold can produce numerous projections with the same dimension as the target distribution on other projection surfaces, with each encapsulating parts of the overarching characteristics of the high-dimensional construct. Therefore, in this paper, we aim to design an operator $f$ that can establish this high-dimensional manifold by estimating the distributions of input and target, and finding the closest projection to the target projection surface. Elaborately, $M = \{(x, z) \in \mathbb{R}^{n_1} \times \mathbb{R}^{n_2} : x \in \{x_i\}_{i=1}^{N}, z \in \{z_i\}_{i=1}^{N} \}$, $f_z: M \rightarrow \mathbb{R}^{n_1}$, we can get the output $f(z)$ as the projection through $f_z$ from the high-dimensional manifold $M$, while $\| f(z) - x \| \approx 0$.

Given data sample of target $\{x_i\}_{i=1}^{N}$, the operator $f$ produces the projection through the high-dimensional manifold space constructed from the inputs $\{z_i\}_{i=1}^{N}$ and $\{x_i\}_{i=1}^{N}$. Additionally, in our design, $f$ is to make this output projection as close as possible to our target distribution $P_x$, and to find the corresponding $x_i'$ that best represent the final output, which is to find two approximative manifolds.

To achieve the objective of approximating the target projection surface, the optimization objective of HDPS is to train the operator $f$ ,which can yield an output as close as possible to the target  $\{x_i\}_{i=1}^{N}$ with the input data $\{z_i\}_{i=1}^{N}$. Assuming a specific situation, when the input data did not meet any distortion, which means that $\{z_i\}_{i=1}^{N} = \{x_i\}_{i=1}^{N}$, we can easily find the most suitable operator $f$. When $f(x) = x$, which can lead to $f(z) = x$, our objective is perfectly fulfilled, and the constructed high-dimensional manifold space refers to the intrinsic spaces of $P_x$ and $P_z$. Therefore, given a distance metric $D$, the drift measure between the output obtained from the input $\{z_i\}_{i=1}^{N}$ and the ideal output can be defined as:
\begin{equation}
\phi_\theta(z)=D(z,f_\theta(z)).
\label{eq5}
\end{equation}
 in which $\theta$ are the parameters of a model $f_\theta$. We hope that the output of the model $f_\theta$ is as close to $x$ as possible when a specific situation occurs, since when  $\{z_i\}_{i=1}^{N} = \{x_i\}_{i=1}^{N}$, we seek to minimize the drift measure:
 \begin{equation}
\min_{\theta}\phi_{\theta}(x) = \min_{\theta}D(x,f_{\theta}(x)).
\label{eq6}
\end{equation}

However, in complex NLOS scenarios, the images captured by the camera often suffer from significant distortion, degradation, and interference caused by scattered noise. Consequently, the distributions of $P_x$ and $P_z$ tend to  differ significantly in NLOS environments. It becomes difficult to establish a direct transformation relationship between $X$ and $Z$ through a single limitation, especially in environments with limited data collection and computational resources. Therefore, our objective is to train the operator \( f_{\theta} \) in such a way that \( z \) evolves into \( f_{\theta}(z) \), and closely resembles \( f_{\theta}(x) \).  In this step, we successfully accomplish  the information integration, which means that the high-dimensional manifold is constructed. Therefore, the ideal scenario involves minimizing the distance between \( f_{\theta}(x) \) and \( f_{\theta}(z) \) while adhering to the constraints specified in Eq.\ref{eq6}. We can obtain the second optimization constraint:
 \begin{equation}
\min_{\theta}\Phi_{\theta}(x,z) = \min_{\theta}D(f_{\theta}(x),f_{\theta}(z)).
\label{eq7}
\end{equation}
\begin{figure}[htbp]%
\centering
\includegraphics[width=0.9\textwidth]{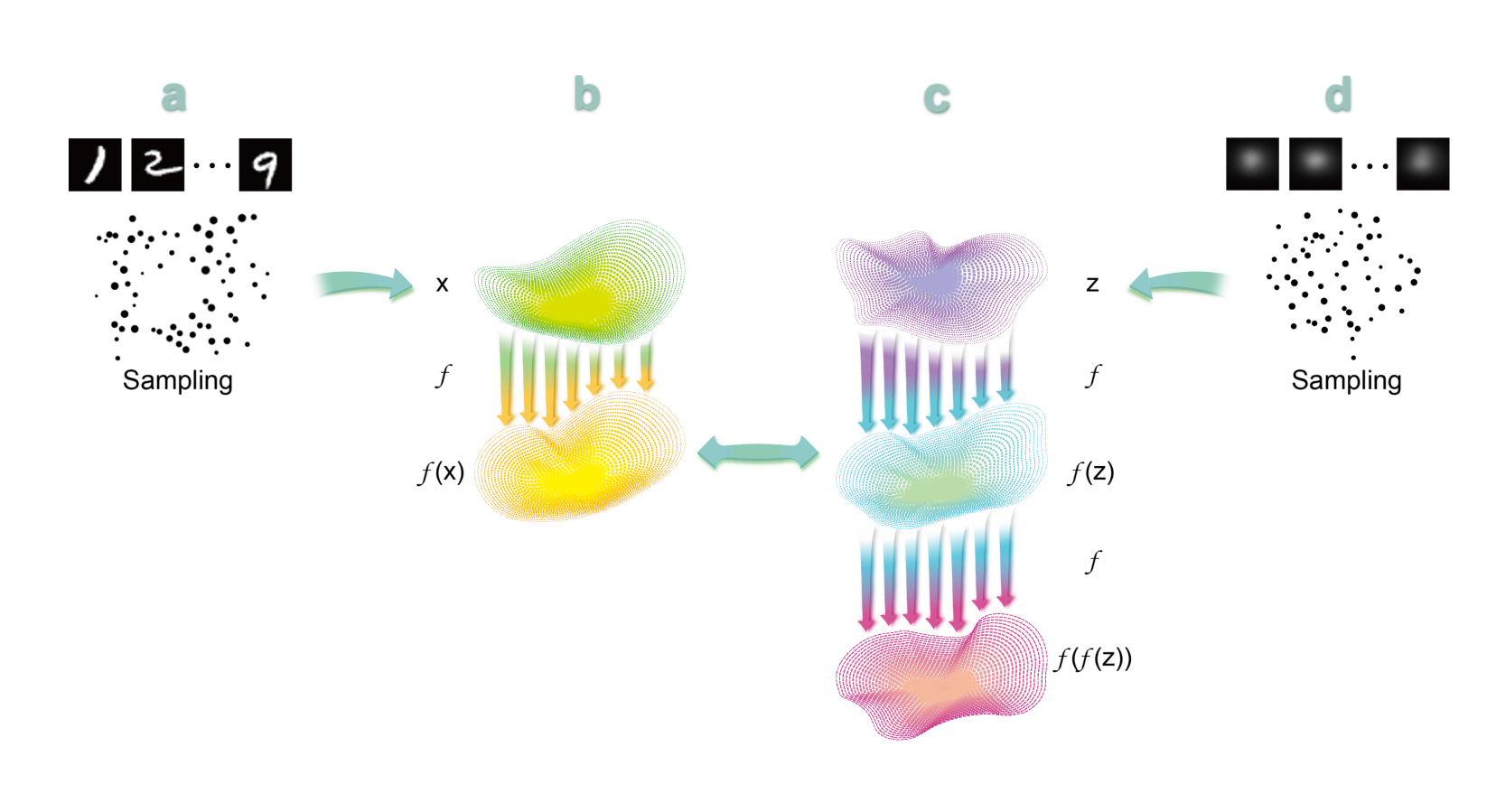}
            \caption{\textbf{The gradient optimization process of HDPS method.} \textbf{a} describes the estimation of the distribution for the target, while \textbf{d} demonstrates the estimation of the distribution for the light spot. \textbf{b} illustrates the process where $x$ is mapped to $fx$ through the operator $f$. \textbf{c} depicts the evolution of $z$ to $f(z)$ through an operator and the process of mapping $f(z)$ as input to $f(f(z))$. The relationship between $f(z)$ and $f(x)$ is established through the high-dimensional manifold constructed by the HDPS method. During the gradient descent process, there are two ways of gradient optimization as shown in \textbf{c}, namely the gradient optimization path of $f$ above and below. To avoid potential conflicts during gradient optimization, we separately train these two paths and freeze the other during training. When the upper path is trained, we aim for $f(z)$ to approach $x$, thereby better manifesting the properties of $x$, striving to achieve the concept of being "sufficiently close" to the target. When the lower path is activated, we aim to decrease the size of the neighborhood of "being sufficiently close".}\label{fig4}
\end{figure}

Certainly, with these two optimization constraints, we partially achieve the task objective. However, to ensure that our output target can be closer to $X$, we hope that the operator $f$ can give $f_z$ some of the properties in $x$.  Based on our previous treatment of \( x \), we observe that we gain minimal alterations by the operator \( f \) when the input appears as the target itself (i.e., when the image is undistorted). Therefore, given that the operator \( f \) satisfies the first two constraints, we can assume that the obtained \( f_z \) is already sufficiently close to \( x \). Hence, we hope that under the influence of the operator \( f \), \( f_z \) also exhibits similar properties, where we can obtain the third optimization constraint:
 \begin{equation}
\min_{\theta}\phi_{\theta}(f_\theta(z)) = \min_{\theta}D(f_{\theta}(z),f_{\theta}(f_{\theta}(z))).
\label{eq8}
\end{equation}
Eq.\ref{eq8} brings a problem, where two $\theta$ appear in the minimized objective and both of which contribute, meaning we can find two distinct pathways of gradients. Therefore, we have to partition it into two components.

To better align the manifold of our output $f(z)$ with the target manifold, we initially optimize only the upper pathway in \textbf{Fig.\ref{fig4}} \textbf{c}, treating the lower one as a frozen state. Given the desire to embody the notion of being "sufficiently close" to the target manifold, we aim for the $f(z)$ obtained through training the upper pathway to reflect characteristics akin to those of the target manifold $\{x\}$, minimizing the distance between $f(z)$ and $f(f(z))$. Hence, our optimization objective is obtained as follows:
 \begin{equation}
\min_{\theta}\phi_{\theta'}(f_\theta(z)) = \min_{\theta}D(f_{\theta}(z),f_{\theta'}(f_{\theta}(z))).
\label{eq9}
\end{equation}
Although Eq.\ref{eq9} encouraged the output $f(z)$ being closer to the target, it is not enough. We are going to tighten the neighborhood of the target manifold to as small as possible, where we only allow a "truly sufficiently close" $f(z)$ to appear. To achieve this goal, we \textbf{maximize} the distance between $f(z)$ and $f(f(z))$. When the upper pathway in Fig.\textbf{\ref{fig4}} \textbf{c} is frozen, we treat $f(z)$ as a given output of the operator $f$ this round. To have a smaller neighborhood of target manifold, we try to treat this $f(z)$ as a defective product, which will encourage f to give a closer $f(z)$ in the next round. This optimizes only the lower pathway, where we can get our optimization objective as follows:
 \begin{equation}
\min_{\theta}-\phi_{\theta}(f_\theta'(z)) = \min_{\theta}-D(f_{\theta'}(z),f_{\theta}(f_{\theta'}(z))).
\label{eq10}
\end{equation}
Combing these above optimizations terms, we get our final term:
\begin{equation}
\begin{aligned}
\mathcal{L}(\theta,\theta') &= \lambda_1\mathcal{L}_{\text{spe}}(\theta) +\lambda_2\mathcal{L}_{\text{cl}}(\theta)+\lambda_3\mathcal{L}_{\text{propX}}(\theta,\theta')+\lambda_4\mathcal{L}_{\text{tight}}(\theta,\theta') \\
&= \mathbb{E}_{x,z}[\lambda_1\phi_{\theta}(x)+\lambda_2\Phi_{\theta}(x,z)+\lambda_3\phi_{\theta}(f_\theta(z))+\lambda_4\phi_{\theta'}(f_\theta(z))].
\end{aligned}
\label{eq11}
\end{equation}
in Eq.\ref{eq11}, $\mathcal{L}_{\text{spe}}(\theta)$ is a particular solution that describes the  property of $x$, and $\mathcal{L}_{\text{cl}}(\theta)$ is the reconstruction term. $\lambda_1$, $\lambda_2$, $\lambda_3$and $\lambda_4$ are the weights of the four parts of HDPS. It is noteworthy that \(\theta\) and \(\theta'\) are numerically equivalent, and the only difference lies in whether numerical updates are conducted during the gradient regression.

Upon establishing the term for HDPS, we can obtain the desired operator $f$ through training. However, within this combination of optimization objectives, we discern that HDPS has two kinds of \"antagonisms\", where operator $f$ is trying to find its minimization with the trade-off between $\mathcal{L}_{\text{spe}}(\theta)$ and  $\mathcal{L}_{\text{cl}}(\theta)$ and the balance of $L_{\text{tight}}$ and $L_{\text{propX}}$. The balance in  $\mathcal{L}_{\text{spe}}(\theta)$ and  $\mathcal{L}_{\text{cl}}(\theta)$ decides whether $f(x)$, the director of $f(z)$ to find its way,  is closer to the target. While the trade-off between $L_{\text{tight}}$ and $L_{\text{propX}}$ lies in the concrete realization of the concept of "sufficient closeness." $L_{\text{tight}}$ aims to negate the current state $f_z$, suggesting that we can achieve a smaller range of target neighborhood manifold. On the other hand, $L_{\text{propX}}$ optimizes whether the current state can better express the property of the target manifold, serving as a default for $f_z$ to be located in the target neighborhood. Balancing these two optimization objectives entails delineating the concept of "how large the target neighborhood range is." With these two antagonisms, when the optimizor meets its minimal term, HDPS will give an improved output to reconstruct the target.

Further, it is worth mentioning that establishing the  HDPS does not depend on any components related to the network structure. Thus, HDPS can be applicable to most networks used for image reconstruction. To avoid potential optimization disputes arising from the introduction of overly complex network structures, we chose the simplest U-net architecture---a fully convolutional U-net structure without inter-layer connections to validate our method.

\section{Result}\label{sec3}
Here we present an overview of the facilities utilized in our experiments, including the equipment, materials, and dataset employed in the NLOS scenes. Following this, we will delve into the performance of the HDPS model, which combines the HDPS method with the foundational U-net network structure, on the NLOS-MNIST dataset that we built and NLOS-passive dataset made by \citet{geng2021passive}.
\begin{figure}[H]%
\centering
\includegraphics[width=0.9\textwidth]{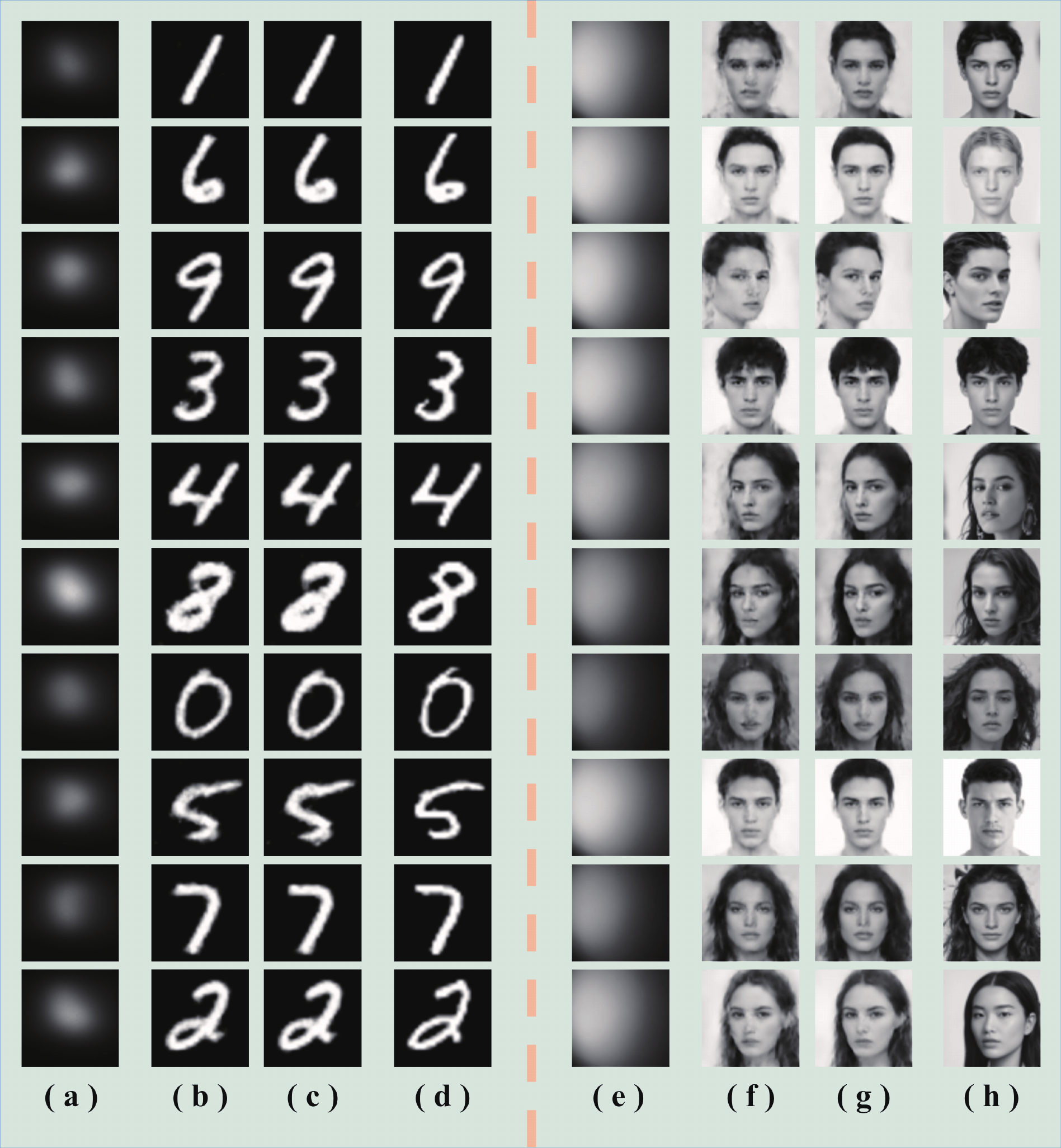}
            \caption{\textbf{Reconstruction of NLOS scenes on two different datasets using HDPS method.} The figure shows experimental results from two datasets. (a) is the Input, and (d) is the target while (b) and (c) are samples $f(z)$and$f(f(z))$ reconstructed with the use of the HDPS method combined with the basic U-Net on the self-assembled NLOS-MNIST dataset. We randomly selected 3000 images from the MNIST dataset as our prior data for training and another 1000 images as unknown data for actual testing. Photographs were conducted on obscured displays in the NLOS environment. The relay surface material chosen for the experiment is of unknown composition, with initial capture angles set arbitrarily to capture speckle information. (e) and (h), which are data from the publicly available NLOS-passive dataset, were used to test and reconstruction, where (e) is the input and (h) is the target. To preserve the characteristics of a small dataset, we randomly selected 3000 images from the extensive NLOS-passive dataset as prior data for training and 1000 images for testing. The presented experimental results, in which (f) is the output $f(z)$ and (g) is $f(f(z))$, are all reconstructions of the test set.}\label{fig5}
\end{figure}

For the NLOS scenes, we employed the Lucid TRI050S-P as our camera, operating in Mono8 mode to capture images of light spots on the relay wall. To restore real-world LOS conditions with unknown material properties and roughness, a black panel, which used to be a part of a discarded computer, was used to simulate the unknown relay wall. The reconstructed scene images were displayed using the SANC-N500-3rd generation, 24-inch IPS 75Hz monitor. The dataset we used is NLOS-MNIST, in which we utilized handwritten digit images from the classic MNIST database as occluded images. Due to our desire for the operator \( f \) trained by our model algorithm to rely on minimal prior knowledge, we employed a minimal number of images as our training set. We randomly selected 3000 images from the MNIST dataset as our training data and additionally selected 1000 images as the test set. Compared to the previous approach utilized 60000 images in Phong's U-Net\cite{zhou2020non}, our demand for prior data was reduced by 95\%. In addition, the Structure Similarity Index Measure (SSIM) of reconstructed images in the test set increased from 0.63 to 0.83, representing a 31.7\% improvement. Furthermore, due to the simplicity of our model's network architecture and the limited amount of training data, we employed a single GeForce GTX 3090 as our training platform. Each training session only consumed 3.6GB of VRAM, resulting in low training costs. To address concerns regarding the generalizability of our method due to its promising performance on a single dataset, we also validated our approach using data from the publicly available NLOS-passive dataset by \citet{geng2021passive}. To maintain experimental conherence, we similarly selected only 3000 images from the face dataset as prior data for training. To ensure consistency in data evaluation, we utilized the author's original open-source testing code for PSNR and calculated, where we obtained our $SSIM = 0.5549, PSNR = 28.47$. The experimental results are shown in Fig.\ref{fig5}.

\begin{figure}[H]%
\centering
\includegraphics[width=0.8\textwidth]{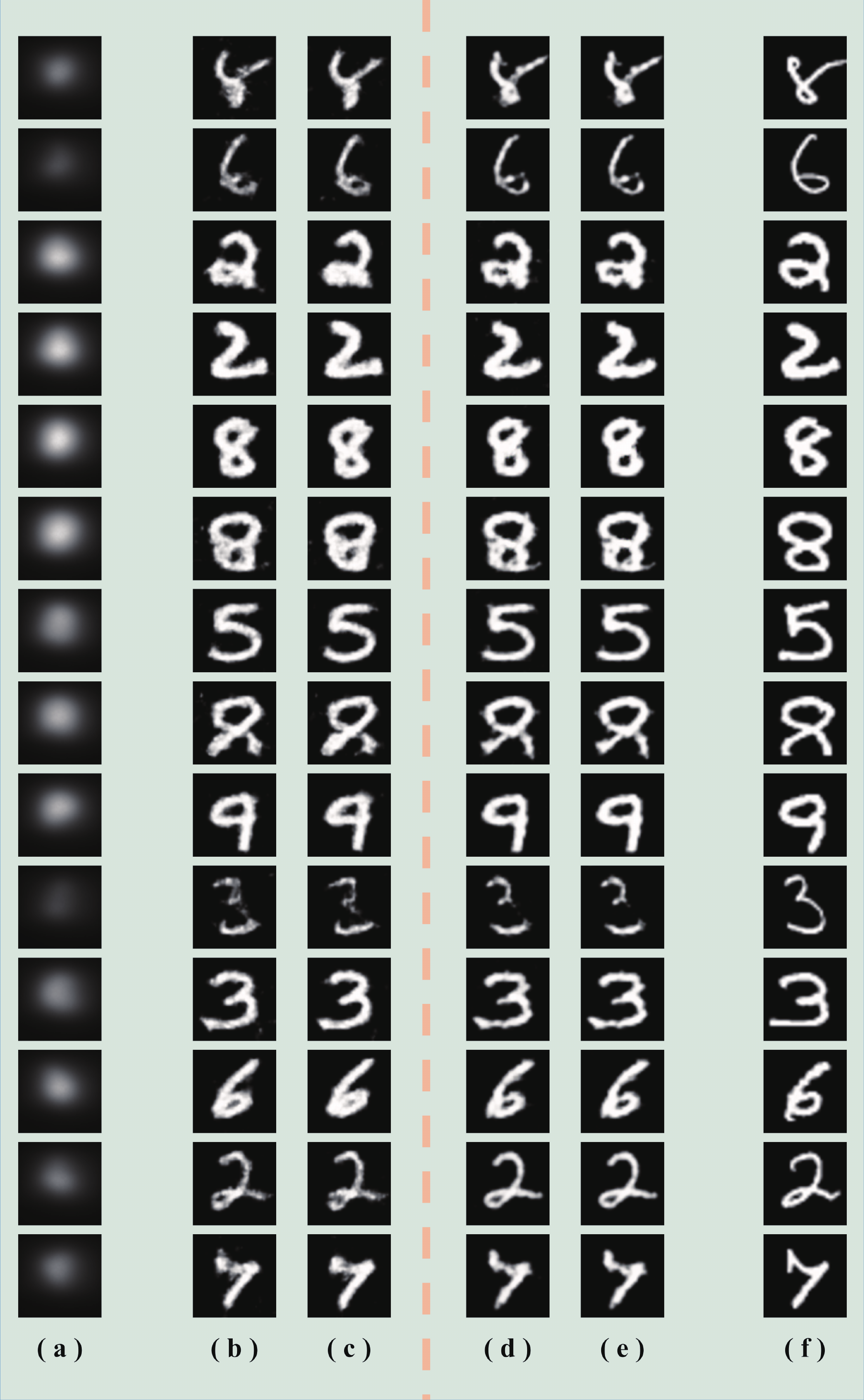}
\caption{\textbf{Comparison of image reconstruction effects using the HDPS framework with different network structures.} The figure shows the performance using different network structures with HDPS. (a) is the input and (f) is the target. (b) and (c) are the reconstruction results $f(z)$ and $f(f(z))$, which used Basic U-Net as the network structure, while (d) and (e) are results using U-Net with skip-connection blocks as the network structure. The output generated by combining HDPS with a superior network structure exhibits stronger reconstruction capability.}
\label{fig6}
\end{figure}

\begin{figure}[H]%
\centering
\includegraphics[width=0.8\textwidth]{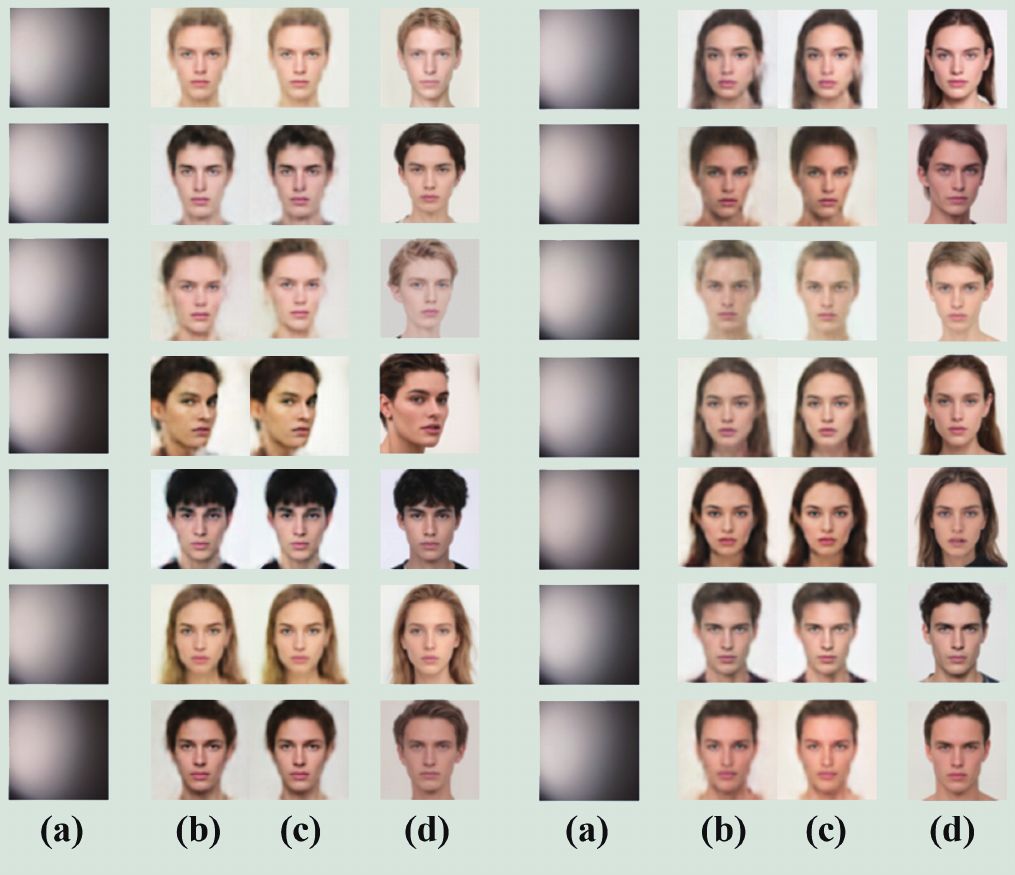}
\caption{\textbf{NLOS reconstruction of RGB images using HDPS framework.} The figure shows the performance reconstructing RGB images. (a) is the input and (d) is the target. (b) and (c) are the reconstruction results $f(z)$ and $f(f(z))$, which used U-Net with skip-connection blocks as the network structure. The experimental results demonstrated the effectiveness of HDPS framework in multichannel tasks.}
\label{fig7}
\end{figure}

The HDPS method achieved better performance in data mapping, fulfilling optimization objectives through a skillful design in mathematical and physical structures. The HDPS framework will enhance the efficiency and capability of reconstruction with optimal combined networks, since HDPS is theoretically network-free. It is mainly used to guide the data flow and mapping in the training of the network. Therefore, the superior the network utilized, the better the data mapping that HDPS will achieve. As such, we substituted the fundamental U-Net structure, integrating skip-connection blocks into it, and conducted experiments on the NLOS-MNIST dataset. The experimental results are shown in Fig.\ref{fig6}. With skip-connection blocks, the SSIM increased from 0.83 to 0.87. In addition, to further demonstrate the effectiveness of our method not only on single-channel gray scale images but also on multi-channel data, we extended our analysis to reconstruct supermodel face data from the NLOS-passive dataset~\cite{geng2021passive} using RGB channels. Employing the same set of 3000 prior data samples, we applied a network architecture with skip-connection blocks to restore supermodel faces in NLOS scenes. On the test set, which is shown in Fig.\ref{fig7}, our method achieved an SSIM of $ 0.6087$ and a PSNR of $28.63$.

The experiment demonstrates that the reconstruction capability of the U-Net structure with skip-connection blocks combined with the HDPS framework surpasses that with Basic U-Net. It exhibits significant improvement in delineating details and interpreting structural information. Table \ref{tab1} indicates the corresponding increases in PSNR and SSIM metrics. Therefore, while HDPS adapts to various network structures, its optimization capability also improves with enhanced network structures.

\begin{table}[h]
\caption{Quantitative comparison of the HDPS framework with different datasets and network models.}\label{tab1}%
\begin{tabular}{@{}lllll@{}}
\toprule
Dataset &network structure & SSIM & PSNR \\
\midrule
NLOS-MNIST    & Basic U-Net   & 0.8302  & 34.41\\
NLOS-MNIST    & U-Net with skip-connection Block  & \textbf{0.8672} & \textbf{34.81}\\
NLOS-passive(Gray)     & Basic U-Net   & 0.5549  & 28.47\\
NLOS-passive(RGB)  & U-Net with skip-connection Block  & \textbf{0.6087} & \textbf{28.63}\\

\botrule
\end{tabular}

\end{table}

\begin{table}[h]
\caption{Quantitative comparison of the HDPS framework with different datasets and network models.}\label{tab1}%
\begin{tabular}{@{}lllll@{}}
\toprule
Dataset &network structure & SSIM & PSNR \\
\midrule
LOL-v1    & U-Net with skip-connection Block   & \textbf{0.898}  & \textbf{28.3}\\
Galdnet from PKU    & U-Net with skip-connection Block  & \textbf{0.8272} & \textbf{28.1}\\

\botrule
\end{tabular}

\end{table}

\section{Discussion}\label{sec4}
We propose a method to obtain passive NLOS scene imaging results with limited prior data. Our model-guided framework design can be adapted to most network structures used for image reconstruction. In this section, we discuss the relationship between HDPS and the Idempotent Generative Networks(IGNs)~\cite{shocher2023idempotent}, another methodology that shares a similar mathematics structure as ours, potential algorithmic improvements for HDPS, and the possibility of explicit expression of the physical and mathematical HDPS model.

We shared our method as a solution for image reconstruction, while IGNs focus on image generation, aiming to generate images from a single step output through the idempotent concept, summarizing the structural properties of the target directly through training an operator. However, for objects with pre-existing objective properties, IGNs may not completely map the manifold they summarize. Moreover, IGNs exhibit weak resolution capability for outputs with significant distribution differences and overlarge targets, and have difficulty in achieving corresponding reconstructions of numerous highly similar samples in the input domain. As the name IGNs suggests, its unique concept of "idempotence" inherently contradicts the losses present in optical transmission processes. 

The conceptual framework of HDPS that we propose constitutes a generalized extension of the mathematical form of IGNs, with a fundamentally different construction approach and with three main differences to IGNs. Firstly, we consider the "idempotence" in IGNs as a special case for situations where the optical propagation distance approaches zero and the relay surface undergoes total reflection. However, such scenarios are impossible practically. Therefore, we introduce a loss term between f(z) and f(x) into the algorithm's structure. This additional loss term not only mitigates the intense conflicts arising from significant distribution differences between the input and target domains but also enables the operator f to spontaneously find the highest upper bound from the projection surface in the input domain to the target projection surface. Secondly, HDPS uniquely introduces the concept of a high-dimensional manifold in its mathematical model. Establishing this high-dimensional manifold provides a feasible training path for the intermediate operator f. Compared to the difficulty of directly building a mapping relationship between two manifolds, estimating the projection of a high-dimensional manifold provides a reasonable explanation for the abstraction of the mapping relationship. Thirdly, HDPS achieves its output by selecting a projection nearest to the target, rather than directly mapping between two manifolds.  During the propagation of light, energy loss inevitably lea
ds to information loss. Moreover, the presence of relay surfaces exacerbates this information loss due to the occurrence of significant scattering. Applying idempotence in such cases is absurd. The substantial loss of information inevitably leads to incomplete estimation of information for this high-dimensional manifold and difficulties in constituting the mapping relationship at once. However, we acknowledge the loss of such information and endeavor to find and ultimately select a minimal neighborhood range in the two forms of confrontation, aiming to identify a projection with minimal information loss as the output of HDPS. 

In Section 2.3, we mentioned that the HDPS method is the result of balancing two antagonistic optimization objectives. Regarding the intermediate quantity $f(x)$, which guides $f(z)$ to achieve better output results, it is influenced by the balance between the distances of x and z. In addition, the optimization objective of imbuing $f(z)$ with the properties of x is another form of balancing objective, emerging from the antagonism between $\mathcal{L}_{\text{propX}}$and $\mathcal{L}_{\text{tight}}$. These two adversarial processes are akin to a tug-of-war game, where the operator $f$ always finds a balancing point between two opposing forces, ensuring that the output results do not lead to reconstruction failures or overfitting.  However, the setting of weights for the four optimization objectives in the HDPS is based on empiric predetermination and cannot be automatically updated from the actual training results. In the future, we hope that the optimization structure we propose will automatically achieve the allocation of weights between various optimization objectives based on macroscopic estimates of the distribution of the target domain and the input domain while ensuring that the gradient descent process does not encounter problems.

Furthermore, from the mathematical perspective, the differences in relay surfaces should not be a reason for the network training operators not to be unified or generalized. From an optical standpoint, differences in relay surfaces mainly arise from variations in roughness and material properties. Roughness significantly affects the scattering phenomena of light after incidence on the relay surface, while materials determine the proportion of light energy loss and the ratio  of reflection and transmission. Both these directly impact the information captured by the receiving camera at the endpoint. However, due to the linear superposition of the light field in NLOS imaging systems, it can be inferred that there must be a one-to-one correspondence between the image information obtained in the target domain under the same relay surface. Additionally, considering the varying degree of light absorption by materials and the complexity of scattering representation, we believe that the representation of the same object with different relay surfaces should also be different, which is merely a result of the variations in projections of high-dimensional manifolds onto different surfaces. Therefore, in the high-dimensional manifold we constructed, the structural information contained in the initial state will not change with variations in projection, meaning that the manifestation of structural information with different relay surfaces should also be consistent in the high-dimensional manifold. In subsequent research, we will endeavor to gradually extend the HDPS method from single relay surfaces to multiple relay surfaces. This will be achieved by adaptively matching the projection direction parameters of operators to complete the imaging reconstruction of NLOS scenes.

In terms of the mathematical model, we have experimentally and theoretically confirmed the existence of SSF and high-dimensional manifolds. In our experiments, we did not use additional optical information such as polarization or wavelength. Rather, we obtained spatial position information, estimated incident vector information, and intensity information only. Therefore, we believe that high-dimensional manifolds are also composed of this optical field information. Deriving from existing information and utilizing Fourier transforms to find their light transport matrices may be a breakthrough direction for the explicit expression of the SSF function. We hope to successfully establish an explicit model for this in subsequent research.

\section{Conclusions}\label{sec5}
In this paper we proposed a few-shot passive NLOS imaging framework, HDPS. By integrating physical model with manifold theory, we constructed a high-dimensional manifold of NLOS data and its projection framework under data and model driven mechanism. With few data, the proposed HDPS method can achieve a stunning accuracy of NLOS imaging reconstruction even with the original Unet as the backbone. To validate the generalizability and robustness of HDPS, we further developed a novel dataset for NLOS imaging, NLOS-MNIST. Extensive experiments are conducted both on NLOS-passive and the proposed NLOS-MNIST datasets. The results demonstrated that our framework is versatile and can be applied to a multitude of network configurations.With the network model refined, the accuracy of HDPS will definitely improve.  



\section*{Supplementary Information}
\subsection*{Acknowledgements}
Not applicable
\subsection*{Authors' contributions} 
Y.T.G, X.M.X and Y.G.F conceived the idea behind the present study. Y.T.G, X.M.X and M.B.Y designed the experiments and collected the data. Y.T.G, X.M.X, Y.G.F and M.B.Y analyzed the results and structured the manuscript. Y.G.F and M.B.Y acquired the funding. All authors contributed to the final version of the manuscript.
\subsection*{Author's information}
Not applicable
\subsection*{Funding}
This work is sponsored by the National Natural Science Foundation of China (NSFC) through grants Nos.62003055 and 62103163.
\subsection*{Availability of data and materials}
The authors declare that the data supporting the findings of this study are available with \url{https://github.com/CVIR-Lab/NLOS}
\section*{Declarations}
\subsection*{Ethics approval and consent to participate}
Not applicable.
\subsection*{Consent for publication}
All authors agree to publish this articles.
\subsection*{Competing interests}
The authors declare no conflict of interest.
\noindent



%
%
%


\end{document}